\documentclass{acmsiggraph} 

\usepackage[scaled=.92]{helvet}
\usepackage{parskip}
\usepackage[labelfont=bf,textfont=it]{caption}
\usepackage[latin1]{inputenc}
\usepackage{graphicx,url}
\usepackage{amsmath} 
\usepackage{color}
\definecolor{grisclair}{rgb}{0.95, 0.95, 0.95}
\definecolor{bleuclair}{rgb}{0.6, 0.8, 1.0}
\usepackage{algorithmic}
\usepackage{algorithm}
\usepackage{amssymb}
\usepackage{amsmath}
\usepackage{pstricks}
\usepackage{colortbl}

\makeatletter
\providecommand*{\toclevel@algorithm}{0}
\makeatother

\title{Video Stippling}
\author{Thomas Houit\thanks{e-mail: thomas.houit@gmail.com}\\ %
\and Frank Nielsen\thanks{e-mail: nielsen@lix.polytechnique.fr}\\}


\begin{document}

\maketitle

\begin{abstract}
In this paper, we consider rendering color videos using a non-photo-realistic art form technique commonly called stippling.
Stippling is the art of rendering images using point sets, possibly with various attributes like sizes, elementary shapes, and colors.
Producing nice stippling is attractive not only for the sake of image depiction but also because it yields a compact vectorial format for storing the semantic information of media.
Moreover, stippling is by construction easily tunable to various device resolutions without 
suffering from bitmap sampling artifacts when resizing. 
The underlying core technique for stippling images is to compute a centroidal Voronoi tessellation on a well-designed underlying density.
This density relates to the image content, and is used to compute a weighted Voronoi diagram. 
By considering videos as image sequences and initializing properly the stippling of one image by the result of its predecessor, one avoids undesirable point flickering artifacts and can produce stippled videos that nevertheless  still exhibit noticeable artifacts. 
To overcome this, our method improves over the naive scheme by considering dynamic point creation and deletion according to the current scene semantic complexity, and show how to 
effectively vectorize video while adjusting for both color and contrast characteristics.
Furthermore, we explain how to produce high quality stippled ``videos'' (eg., fully dynamic spatio-temporal point sets) for media containing various fading effects, like quick motions of objects or progressive shot changes.
We report on practical performances of our implementation, and present several stippled video results rendered on-the-fly using our viewer that allows both spatio-temporal dynamic rescaling (eg., upscale vectorially frame rate).
\end{abstract}

\keywordlist

\section{Introduction}

\copyrightspace

\begin{figure}
\centering
\includegraphics[width=0.45\columnwidth]{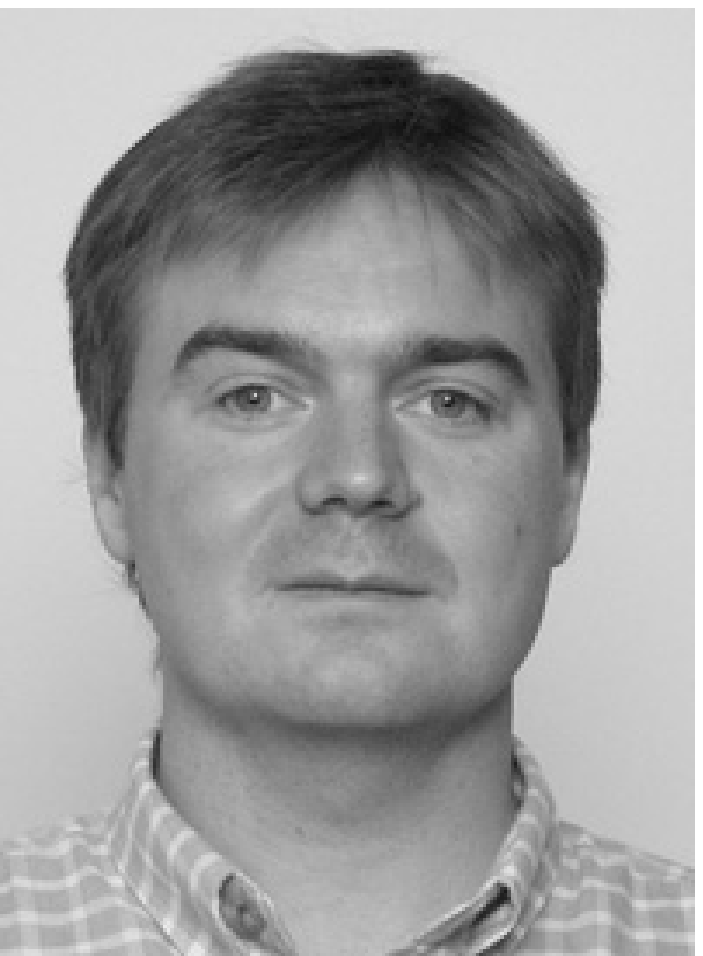}
\includegraphics[width=0.45\columnwidth]{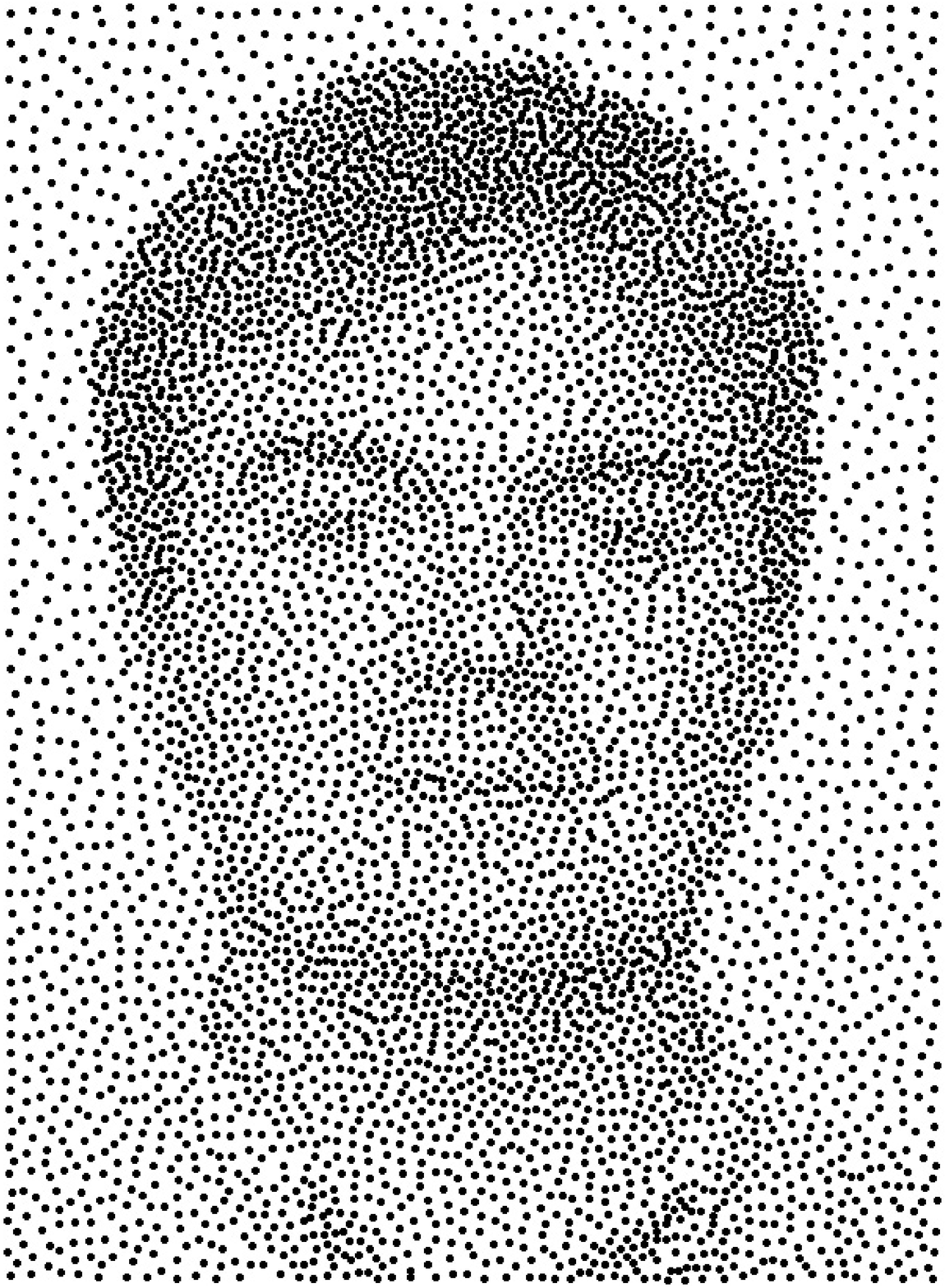}
\caption{Stippling a photo by rendering the bitmap (left) as a point set (right) distributed uniformly according to a prescribed underlying image density: the grey-level intensity (here, 600 points).\label{fig:frank}}
\label{frank}
\end{figure}

Historically, stippling was primarily used in the printing industry to create dot patterns with various shade and size effects. 
This technique has proven successful to convey visual information, and was widely adopted by artists that called this rendering art {\it pointillism}.\footnote{See~\url{http://www.randyglassstudio.com/} for some renderings by award-winning artist Randy Glass.} 
Informally speaking, the main idea is that many dots carefully drawn on a paper can fairly approximate different tones perceived by local differences of density, as exemplified in Figure~\ref{frank};
 To the Human eyes, a high density region looks darker than a low density one.
 The main difference with dithering and half-toning methods is that points are allowed to be placed anywhere, and not only on a fixed regular grid.
  Thus the primary difficulty in stippling is to obtain a point distribution that adapts to a given density function, in the sense that the number of points in an area must be proportional to the underlying density.

\medskip

In this work, we present a method to generate stippled videos. Generating videos with stipples or other marks is very interesting but has not yet be fully solved.
Our method produces high quality videos without point flickering artifacts. 
Each point can easily be tracked during an entire shot sequence. 
Our algorithm is able to render fading effects\footnote{Fading effects are so commonly met in practice in videos following a storytelling that they cannot be neglected.} with the adaptation of both point and color densities. 
Our method allows us to deal robustly with an entire video without having to cut it down into pieces using shot detection algorithms. 
To be able to work on the ``output video'' (time-varying point sets) and adjust dynamically some rendering parameters, we also developed an in-house video viewer that handles these vectorized stippled video. 
The viewer let us choose the contrast to adjust point section, let us use color codings or introduce patterns for drawing edges to improve the rendering quality.
Moreover, the viewer application allows one to interpolate unknown frames at any time position to match the frame rate of the display device.
Thus, we can easily increase the frame rate if required, or conversely slow down some portions of the stippled videos. To improve the edge render the viewer application offers one to replace points by patterns oriented with the local gradient of the current frame.

\medskip
Secord~\cite{DBLP:conf/npar/Secord02} designed a non-interactive technique to create quality stippling still images in 2002. 
Based on the celebrated Lloyd's $k$-means method~\cite{DBLP:journals/tit/Lloyd82}, Secord creates stippling adapted to a density function and improves the result by considering various point sizes.We considered this approach of Stippling for our algorithm because it fits with the goal of our algorithm: being able to render an high quality output for every type of video without knowing any other data. There are other approaches like the use of multi-agent systems to distributes stipples has been explored by~\cite{DBLP:journals/cgf/SchlechtwegGS05}. This way of rendering is easily extendable to various kind of stroke-based render. There is too a possibility of generating stipple patterns on 3D-objects by triangulating the surface and by positioning the dots on the triangle vertices~\cite{DBLP:journals/cga/PastorFS03}. This is a very fast method but the resulting patterns are not optimal. This has been improved by~\cite{VBTS07a} but we always cannot use as input every type of data --- such as an AVI file taken on the internet. 

\medskip
Like all computer-generated point distributions,  we are also looking for stippled images that has the so-called {\it blue noise}  characteristics with large mutual inter-distances between points and no apparent regularity artifacts. An interesting approach of stippling with blue noise characteristics based on Penrose tiling is described by Ostromoukhov~\cite{DBLP:journals/tog/OstromoukhovDJ04}. More recently a method based on Wang tiles enabled the creation of real-time stippling with blue noise characteristics. This method was described by~\cite{KCODL06}. Balzer et al.~\cite{DBLP:journals/tog/BalzerSD09} developed another alternative way to create, this time Voronoi diagrams with blue noise characteristics. They introduced so-called capacity constrained diagrams that converge towards distributions that exhibit no regularity artifacts and which adaptation to given heterogeneous density function behaves better. 
This capacity-constrained method enhances the spectral properties of point distribution while avoiding its drawbacks. Our algorithm can be used based on both Secord of Balzer et al. approaches and inherit of their blue noise characteristics. We choose the Secord algorithm for rapidity, or the Balzer et al. algorithm for high quality point distribution.

\medskip
Another way to render a stipple drawings is to mix different type of points instead of only use contrasted and coloured points. The work of S. Hiller~\cite{DBLP:journals/cgf/HillerHD03} explored this possibility by positioning small patterns in a visually pleasing form in the stipple image. We use in our algorithm patterns to increase the render of the edges.

\medskip
We adapted these recent set of methods to video and solved along problems that appeared while doing so. 
A major inconvenient of these previous stippling methods was that while those methods can handle nicely and efficiently fine details and textures, they
fall short when dealing with objects with sharp edges. 
In this paper, we overcome this drawback and propose a frequency-based approach in order to detect images edges and enhanced their support and rendering. 
This frequency approach is improved by the use of patterns that replace some points and partially reconstruct the edges of the image. It is another way to use the frequency approach to improve the rendering of edges and this without increasing their size, but rather by suggesting their local orientation according to the surround shape elements~\cite{DBLP:conf/psivt/GomesSC07}.

\medskip

The roadmap of the paper is as follows:
The basic ingredient of our approach is the use of {\it centroidal Voronoi diagrams} (a fundamental structure of computational geometry) to produce good-looking distributions of points, as explained in Section~\ref{voronoi}. 
This generative method will be further extended to compute a full video in Section~\ref{video}. 
Then to improve the rendering, one need to change some parameters to obtain good fading effects, as explained in Section~\ref{diff}. An improved rendering must also consider both color information and point contrasts. Small adjustments described in Section~\ref{adjust} are required to get such a desirable output. 
We implemented a tailored scheme to handle the stippling rendering of sharp edges. 
Section~\ref{frequence} explains our solution to improve edge rendering by considering both high and low frequencies to place  points accordingly.
Finally we used some oriented patterns to reconstruct sharp edges in the output.
Section~\ref{pattern} explains the method used and the describes the obtained results.

\section{Voronoi diagrams}
\label{voronoi}
In order to create a stipple image, we adapted Lloyd's $k$-means method~\cite{OkabeBootsSugihara} to obtain a Voronoi diagram that adapts to an underlying image density. To do this, we need two sets of points:
\begin{description}
\item The first point set which is called the {\it support set} of points is chosen following the grey-scale intensity density of the image.
 In fact, we choose in practice $N$ points in the source image, where $N$ can be larger than the total number of pixels in the image. 
Those support points should be drawn such that if a pixel is darker compared to another one, that pixel should have more chance to contain one or more support points. 
To get those support points, we choose one point coordinate $(x,y)$ at random, and then an integer falling in the grey intensity range $[0,255]$. 
If this integer is greater than $255$ minus the color of the pixel at $(x,y)$, we decide to put a support point in $(x,y)$.

\item Then we generate the second set of points which will create the Voronoi diagram itself. 
We call these points: \textit{Sites}. 
This second set of points has the same characteristics as the support points. Only the number of points changes;
 We choose only $N/\alpha$ points --- were $\alpha = 10^{3}$ for example.
\end{description}

\medskip
Then every generator point is identified with the closest site according to a given metric. 
The common choice is to use the Euclidean $L^{2}$ distance metric 

\begin{equation*}
d(P_1,P_2)=\| P_{1}-P_{2} \| = \sqrt{(x_{1}-x_{2})^{2} + (y_{1}-y_{2})^{2}}
\end{equation*}

where $P_{i} = (x_{i},y_{i})$ are two given points. 
Each set of points identified with a particular site forms a Voronoi region, and the set of Voronoi regions ``covers'' the entire image. 
We can define each Voronoi region $V_{i}$ as follows:

\begin{multline*}
V_{i} = \Bigl\{ x\in \Omega \ |  \  \| x-x_{i}\| \leq \| x-x_{j}\|  \\
\mathrm{for} \  j=1,...,n \  \mathrm{and} \  j\neq i \Bigr\}
\end{multline*}

where $\|.\|$ denotes the Euclidean distance derived from the $L^{2}$ norm.

\medskip
The following update step is performed to obtain a good-looking random distribution of sites:
We move each site to the mass center of its associated generator points. 
Given a density function $\rho (x) \geq 0$ defined on $\Omega$, for any region $V \subset \Omega$, the standard mass center $C$ of $V$ is given by:

\begin{equation*}
C = \frac{\displaystyle\int_{V} x \rho(x)dx}{\displaystyle\int_{V} \rho(x)dx}
\end{equation*}

Here, we consider $\rho(x) = 1$. Then we do again the identification for each generator point.  Iterating these two steps let us converge to a Central Voronoi diagram~\cite{CVT}, a quasi-uniform partition of the domain adapted to a given image density. 
The CVT algorithm operations are summarized in Algorithm~1.

\begin{algorithm}
\caption{Centroidal Voronoi Tesselation (CVT).}
\begin{algorithmic}[1]
\STATE \underline{Generation of support points}:
\FOR{$i = 1$ to Number of Sites $*\ \alpha$}
	\STATE $x = $ Random Integer $\in [0;\mathrm{Width}]$
	\STATE $y = $ Random Integer $\in [0;\mathrm{Height}]$
	\STATE $m = $ Random Integer $\in [0;255]$
	\IF {$m > 255-$Color$(x,y)$}
		\STATE Add support point at $(x,y)$
		\STATE $i \leftarrow i + 1$
	\ENDIF
\ENDFOR
\STATE \underline{Generation of sites}:
\FOR{$i = 1$ to Number of Sites}
	\STATE $x = $ Random Integer $\in [0;\mathrm{Width}]$
	\STATE $y = $ Random Integer $\in [0;\mathrm{Height}]$
	\STATE $m = $ Random Integer $\in [0;255]$
	\IF {$m > 255-$Color$(x,y)$}
		\STATE Add Site at $(x,y)$
		\STATE $i \leftarrow i + 1$
	\ENDIF
\ENDFOR
\STATE \underline{Associate support points to closest sites}
\FOR{$i = 1$ to Number of Sites $*\ \alpha$}
	\STATE Find the closest Site to the support point $i$
	\STATE $d$ Smallest distance
	\STATE $id$ Id of the nearest Site
	\FOR{$j = 1$ to Number of Sites}
		\STATE Calculate the distance $l$ between Site $j$ and support point $i$
		\IF {$l < d$}
			\STATE $d = l$
			\STATE $id = j$
		\ENDIF
	\ENDFOR
	\STATE Associate support point $i$ with Site $id$
\ENDFOR
\WHILE {Convergence criteria $ < $ threshold}
	\STATE Move sites to mass center
\ENDWHILE

\end{algorithmic}
\end{algorithm}

\begin{figure}
\centering
\includegraphics[width=0.38\textwidth]{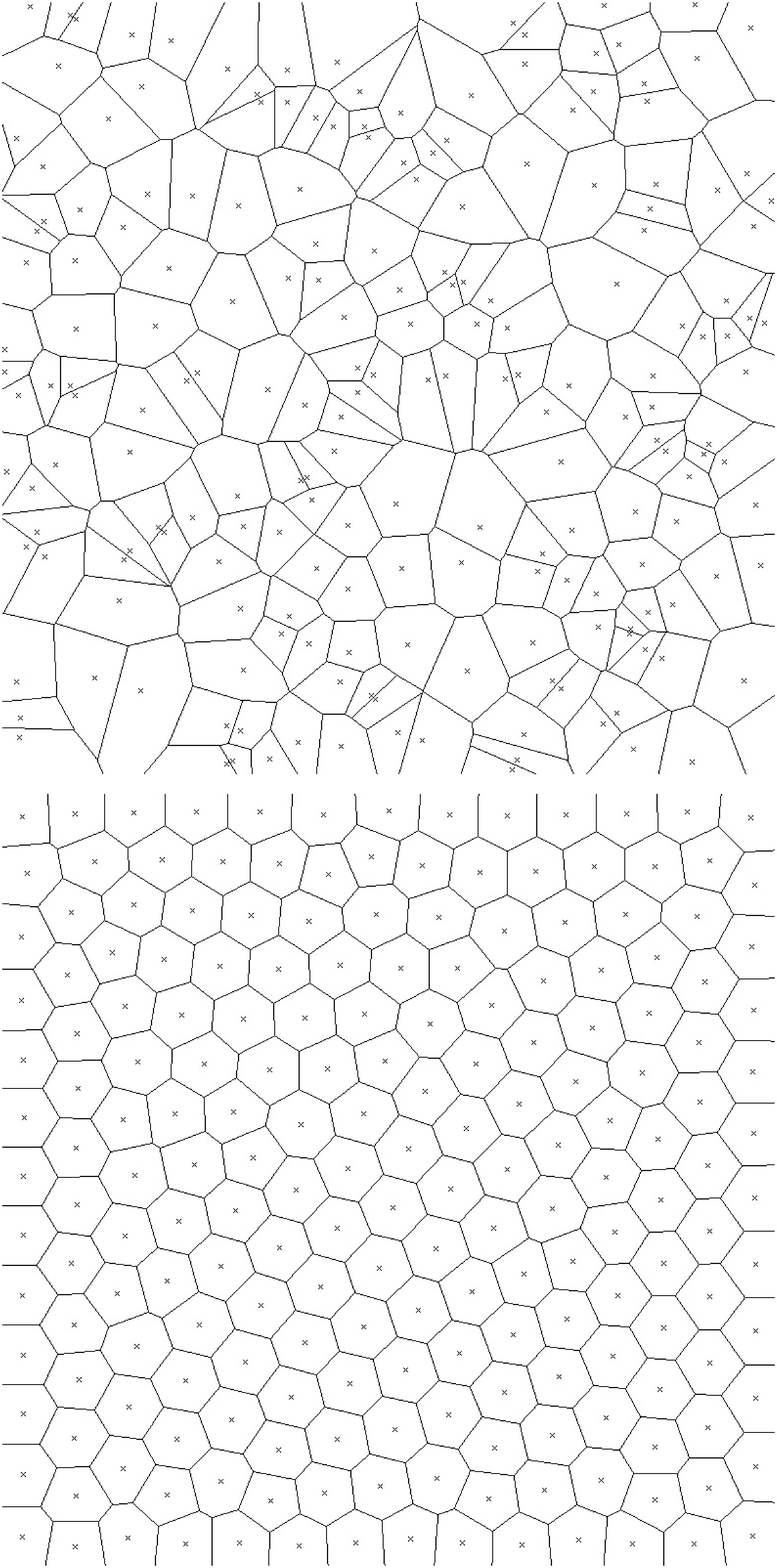}
\caption{Centroidal Voronoi Diagram (bottom)  generated from  successive Voronoi diagrams (top). 
Each site is iteratively relocated to the centroid (the  center of mass) of its Voronoi region.}
\label{figVoronoi}
\end{figure}

\medskip
The convergence of Lloyd's algorithm to a centroidal Voronoi diagram on continuous domains has been proven for the one-dimensional case. 
Although the higher dimensional cases seem to converge similarly in practice, no formal proof is yet reported. 
Notice that here we fully discretize the space by considering support points.
The criterion we used to define the convergence is the numeric comparison between the positions of each site between two consecutive iterations. 
If the sum of those distances between the two positions of each site during the last iteration falls under a prescribed threshold, we consider that the computation is terminated. 
This discretized version of CVT provides good and quick results as attested in~Figure~\ref{figVoronoi}.

\medskip
Another common solution is to stop Lloyd's relocation scheme after a prescribed number of iterations. 
Lagae~\cite{DBLP:journals/cgf/LagaeD08} suggested to use the normalized Poisson disk radius $\alpha \in [0;1]$ as a quality measure for point distributions. If two points in the distribution coincide, $\alpha = 0$. If there is a hexagonal lattice, $\alpha = 1$ and Lagae~\cite{DBLP:journals/cgf/LagaeD08} chose $\alpha = 0.75$ as optimal for a reference point set obtained via dart throwing (a common rejection sampling method). The convergence of $\alpha$ can be utilized as a termination criterion, where we stop Lloyd's method as soon as $\alpha$ becomes stable.

\section{CVT method for video stippling}

\subsection{From images to video}
\label{video}
The first step consists in computing the first image stippling as explained in Section~\ref{voronoi}. 
We store the image and the position of each generator point and each site in a text file. 
The information we need for the remaining of the algorithm is the current density, the current number of generator points and the current number of sites. We need also to be able to identify each site. 
That is why we keep in the text file the identification number of all sites. 
This way of storing data gives us the possibility to pause our algorithm if needed, and resume it later.

\medskip
To compute a full video stippling, we first extract from the video an image sequence in order to be able to use image stippling methods. We then need to keep the information about stipple points on the $N-1$ images before computing the $N$-th image. If we fail to doing so, no correlation between the images of the sequence will appear and undesirable buffering effects will appear in the synthesized stipple video. 
We start the computation of the $N$-th image with the relocation of generator points and sites found for the $(N-1)$-th image by reading the corresponding text file. Then we seek for all the generator points and sites that are no longer needed in this new image. 
To decide whether to keep points or not, we generate a {\it difference image} where each pixel intensity is set as follows:
 
\begin{equation*}
\left\{
  \begin{aligned}
    P_{\mathrm{diff}}(x,y) & = 0 \text{ if } P_{N-1}(x,y)-P_{N}(x,y) \leq 0 \\
    P_{\mathrm{diff}}(x,y) & = P_{N-1}(x,y)-P_{N}(x,y) \text{, otherwise}\\
  \end{aligned}
\right.
\end{equation*}

where $P_{N}(x,y)$ is the color of the pixel at the coordinates $(x,y)$ of the image frame number $N$.

\medskip
Each generator point or site placed on a pixel where the color of the difference image is different from $0$ has a probability  proportional to the value of the color to be deleted. 
If the value is $255$ the probability is $100\%$. We do the same to find where we have to {\it add} new points by calculating another difference image  as follows:

\begin{equation*}
\left\{
  \begin{aligned}
    P_{\mathrm{diff}}(x,y) & = 0 \text{ if } P_{N}(x,y)-P_{N-1}(x,y) \leq 0 \\
    P_{\mathrm{diff}}(x,y) & = P_{N}(x,y)-P_{N-1}(x,y) \text{, otherwise}\\
  \end{aligned}
\right.
\end{equation*}

\begin{figure*}[ht]
\centering
\includegraphics[width=\textwidth]{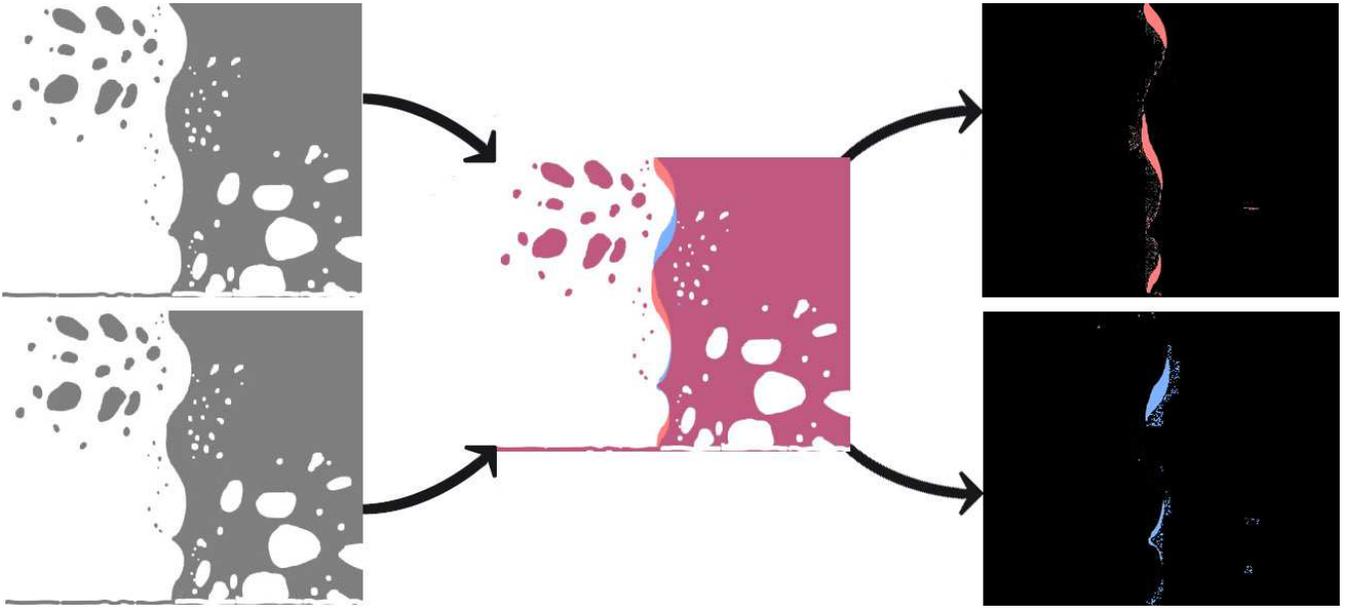}
\caption{From two consecutive frames (left), we generate two difference images (right). 
The evolution between the  first two frames is shown on the central image.}
\label{figImgdiff}
\end{figure*}

We obtain two difference images as shown in Figure~\ref{figImgdiff}.

\medskip
Then we need to link the global density of the image with the total number of points. Thus we calculate for each image its {\it global density} which is equal to 

\begin{equation*}
\text{Density } d = \frac{\sum_{i=1}^n P(i)}{255 n}
\end{equation*}

where $P(i)$ is the grey color of the pixel $i$ and $n$ is the total number of pixel. The user enter the initial number of points. This number of point is linked to the initial density and serves as a {\it reference} during the processing of the full video sequence. 
Our algorithm preserves the same ratio between the number of points and the image density during all the operations. 

\medskip
We repeat this step for all the images of the video sequence. When it is done, we need to produce a final file containing all relevant information. Our program implementing the above stippling method reads all text files calculated during the process and store the evolution of each site identified by its own unique number. For each site we store at each frame its current position and color. 
With this information we generate a text file where we also add other important information such as the total number of sites that will appear during the whole shot, the size of the video (eg., width and height dimensions), and the total number of frames. 

\medskip
To read and ``play'' the output file, we have developed an in-house application in Java\texttrademark{} that renders the stippled video. 
With the information provided in the text file, the viewer application let us resize dynamically the video on-demand, change the contrast of the sites (the difference of size between a black site and a white one) and allows users to activate or not the color mode. 
Moreover, our application allows one to generate intermediate time frames by interpolating the position, size, and color of each site between two consecutive frames. Thus yielding true spatio-temporal vectorization of video media.

The operation work flow carried out by the viewer to visualize stippled videos are explained in Figure~\ref{viewer}.

\begin{figure}
\fcolorbox{grisclair}{grisclair}{
\begin{minipage}[c]{0.45\textwidth}
Read output file:

\begin{minipage}[c]{0.95\textwidth}
\centering
\footnotesize
\begin{tabular}{|c|c|c|c|c|c|}
\hline  & \multicolumn{3}{c|}{Frame 1} & \multicolumn{2}{c|}{Frame 2} \\
\hline Site ID & x & y & RGB Color & x & ... \\ 
\hline 1 &  &  &  &  &  \\ 
\hline 2 &  &  &  &  &  \\ 
\hline ... &  &  &  &  &  \\ 
\hline 
\end{tabular} 
\end{minipage}

\begin{minipage}[c]{0.95\textwidth}
\centering
\ifx\JPicScale\undefined\def\JPicScale{1}\fi
\psset{unit=\JPicScale mm}
\psset{linewidth=0.3,dotsep=1,hatchwidth=0.3,hatchsep=1.5,shadowsize=1,dimen=middle}
\psset{dotsize=0.7 2.5,dotscale=1 1,fillcolor=black}
\psset{arrowsize=1 2,arrowlength=1,arrowinset=0.25,tbarsize=0.7 5,bracketlength=0.15,rbracketlength=0.15}
\begin{pspicture}(0,0)(0,15)
\newrgbcolor{userLinecolor}{0 0.8 0.8}
\psline[linewidth=3,linecolor=userLinecolor]{->}(0,15)(0,0)
\end{pspicture}
\end{minipage}

\begin{minipage}[c]{0.95\textwidth}
Draw points with these characteristics:
\begin{equation*}
\footnotesize
\begin{aligned}
N & = \mathrm{Frame Number} \\
\mathrm{Weight} & = N - \lfloor N \rfloor \\
x & = x_{\lfloor N \rfloor}*(1-\mathrm{Weight})  + x_{\lceil N \rceil}*\mathrm{Weight} \\
y & = y_{\lfloor N \rfloor}*(1-\mathrm{Weight})  + y_{\lceil N \rceil}*\mathrm{Weight} \\
\mathrm{Color} & = \mathrm{Color}_{\lfloor N \rfloor}*(1-\mathrm{Weight})  + \mathrm{Color}_{\lceil N \rceil}*\mathrm{Weight} \\
\mathrm{Size} & = \frac{\mathrm{Color}}{255} * \mathrm{Contrast} \\
\end{aligned}
\end{equation*}
\end{minipage}
\end{minipage}
}
\caption{The Java viewer first reads the source file, then renders on-the-fly the required frame by interpolating the frame and its characteristics from two consecutive discrete time frames.}
\label{viewer}
\end{figure}

\subsection{Image differences and  drift adjustment of sites}
\label{diff}
\subsubsection{Handling fading effects}

The method described before is particularly well adapted to shots where objects appear and disappear instantly. The problem is that there are often objects that disappear using a fading effect. 
If we use the previous formula to calculate the image differences,
 we do not obtain a proper rendering for those fading transition shots. 
 Indeed, if an object disappears progressively during three images (saym $33\%$ each time), we suppress $33\%$ of the generator points and site at each time. But ${\left({\left(100*0.33\right)}*0.33\right)}*0.33 > 0$. So there are generator points and sites in white zones of the image after the  full removal of the object, which is of course not the expected result. 
 To correct this step, we needed to find a difference formula that converges towards $0$ when the destination image becomes fully white. 
 We implemented in our program the following formula:

\medskip
If $P_{N-1}(i)<P_{N}(i)$ we store in the difference image:
\begin{equation*}
{\Bigl(P_{N}(i)-P_{N-1}(i)\Bigr)}\times\frac{255}{255-P_{N-1}(i)}
\end{equation*}

If $P_{N-1}(i)>P_{N}(i)$ we store in the other difference image:
\begin{equation*}
{\Bigl(P_{N-1}(i) - P_{N}(i)\Bigr)}\times\frac{255}{P_{N-1}(i)}
\end{equation*}
using the same notation as before.

\begin{figure*}[ht]
\centering
\includegraphics[width=\textwidth]{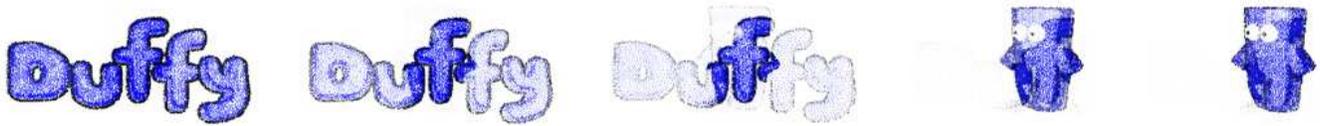}
\caption{The text ``Duffy'' (left) fades out progressively to let the small character appear (right). The point density of Voronoi tesselations adapts automatically and progressively to this fading effect. (3000 points + 1500 frequency points)}
\label{figImgfading}
\end{figure*}

\medskip
This yields a perfect image/vector transcoding of the fading effect during a shot sequence. 
A work-out example is shown on figure \ref{figImgfading}.

\subsubsection{Drift correction}

Another major drawback observed is the drift of the sites which are on an edge of an object in the image. After putting the sites on the mass center of the associated generator points, it appears that the sites are slightly located on the wrong side of the boundary, often on a white zone. 
To avoid this effect, we had to add another step in our algorithm. It is needed to do another step for removal and addition of sites and generator points to clean the image. Our algorithm is reported in Algorithm~2.

\begin{algorithm}
\caption{Next image computation algorithm}
\begin{algorithmic}[2]
\STATE Difference images calculation
\STATE New density calculation
\STATE Total number of points and Sites update
\FOR{$i = 1$ to 2} 
	\STATE $i \leftarrow i + 1$
	\STATE Support points and Sites suppression
	\WHILE {Number of Sites $< $ Total number of Site}
		\STATE Add Site
	\ENDWHILE
	\WHILE {Number of support points $< $ Total number of points}
		\STATE Add support point
	\ENDWHILE
	\STATE Move Sites to mass center
\ENDFOR
\WHILE {Convergence criteria $ < $ threshold}
	\STATE Move Sites to mass center
\ENDWHILE

\end{algorithmic}
\end{algorithm}

\begin{figure}[ht]
\centering
\includegraphics[width=0.48\textwidth]{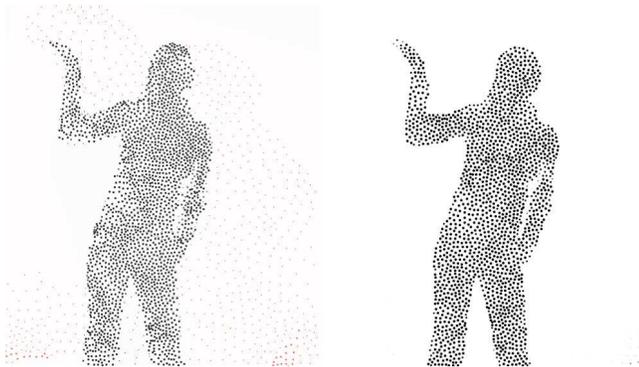}
\caption{(Left) Some points drift from the original shape because of their identification with the centroid of their Voronoi region. 
(Right) our method removed this undesirable effect. (2000 points)}
\label{imgDev}
\end{figure}

\medskip
To completely suppress this annoying artefact, we need to repeat the operation of suppression and the relocation of the sites several times until the number of  deleted sites falls under a threshold. 
In practice, we noticed that repeating this step only one more time is already very efficient. 
The result is shown in~Figure~\ref{imgDev}.

\subsection{Color and contrast}
\label{adjust}
It would seem that to faithfully represent a grey-scale image of 256 levels, our method would require 256 support points per pixel of the original image. This is fortunately not the case. Indeed when rendering a stippling image we lose some information by the lack of support point. But we work on ``areas'' and not on ``pixels'' with stippling method. The most important information is the tone of each area of the image. The exact representation of the support of a stipple image is rather a segmented image without border and with gradient filled areas than a bunch of pixels. And in order to save the lost of ``local'' information we can easily improve the rendering by considering some other options. That is why we consider first the colour and the variation of the size of respective point sites. 

\begin{figure}
\centering
\includegraphics[width=0.48\textwidth]{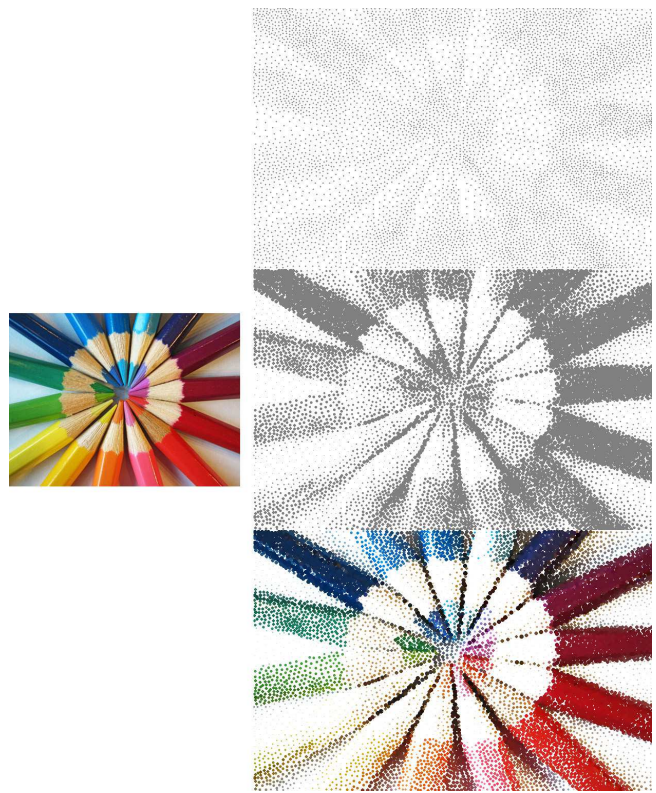}
\caption{(Top) Stippling result without contrast and color. (Middle) stippling with contrast. (Bottom) stippling with color to enhance the rendering. (3000 points)}
\label{greytocolor}
\end{figure}

\medskip
To each image, we can further add some properties to each site. It is easy to pick up the color of the pixel of the reference image which is at the same position of the site. With this information we can display the site in it original color and adapt its size in proportion to its grey color. A white site will be very small and a black site will have the biggest size. This simple operation improves considerably the perception of the  rendering of the output stippled ``video'' and let the user distinguish more details in videos with low contrasts (see Figure~\ref{greytocolor}).

\subsection{Frequency consideration}
\label{frequence}
\begin{figure}
\centering
\includegraphics[width = 0.38\textwidth]{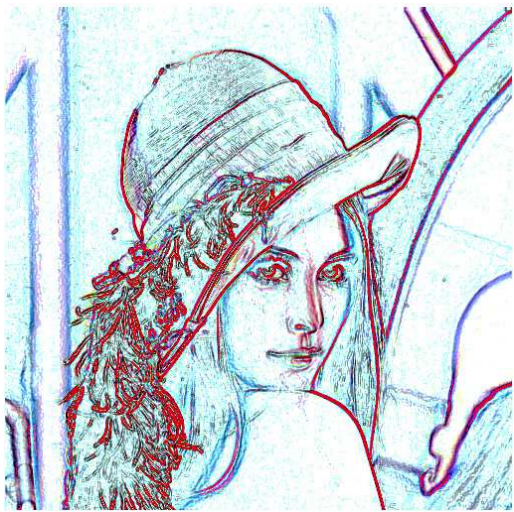}
\caption{Discrete gradient Sobel operator on the image with a threshold that allows one to remove minor details.}
\label{lenaSobel}
\end{figure}

In order to detect edges and enhance their support and rendering we considered a novel rendering approach. 
This operation is carried out in parallel of the previous video processing. 
We first compute the discrete gradient at all pixels the images (using the Sobel operator) and filter the result with a threshold to put low frequencies at zero, as depicted in Figure~\ref{lenaSobel}. Then we apply the same algorithm with a number of site that can vary and that has to be adapted considering the number of pixels with high frequencies. For instance for a color image without much contrast like {\tt Lena}, we have to add twice more points than the original number to have a nice render. But this has to be adapted by the user empirically.

\begin{figure*}
\centering

\includegraphics[width=\textwidth]{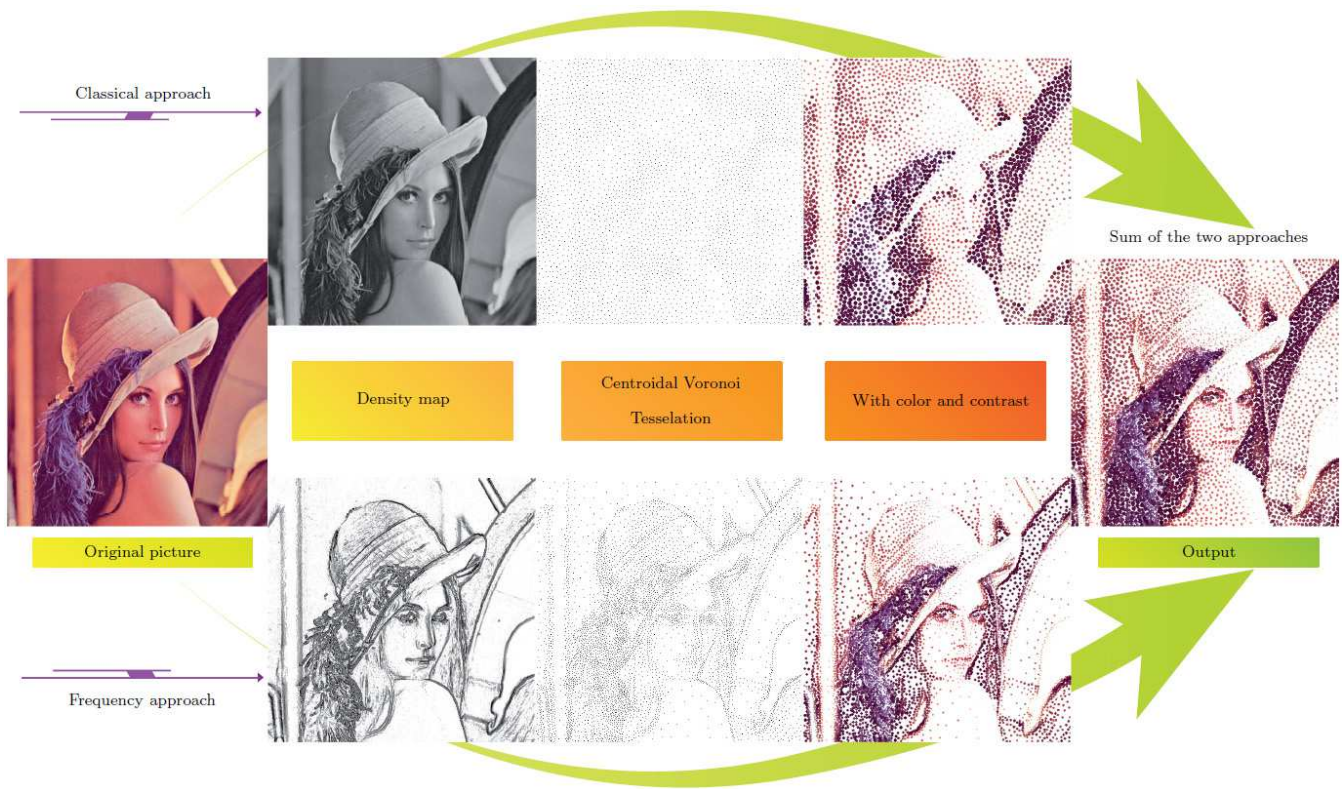}
\caption{We extract two density maps from the original source image: one is the color map and the other the frequency map. Then we apply our stippling algorithm and finally add both contrast and color information. We end the process by summing these two contributions --- the classical (3000 points) and frequency approaches (6000 points).}
\label{lena}
\end{figure*}

\medskip
We obtain a distribution of sites located on the edges of the image. 
These edge points let us improve the overall rendering of the shapes in the image. 
To get a good rendering, we merely add those sites to the sites previously calculated. 
To get a better rendering we reduce the size of these points by 33\% in comparison to the other points. 
The frequency-based stippling result is presented for {\tt Lena} in the Figure~\ref{lena}.

\subsection{Pattern placement}
\label{pattern}

Detecting edges is easy but rendering them clearly by stippling is not really possible with a small amount of Sites. In order to keep a good render without having to generate a lot of Sites, we considered a way to replace the points used to represent the Sites previously obtained by the frequency approach, by patterns. The best solution is to use small segments that follow the edge. Thus, with a small quantity of segments we are able to recreate the edges.

\medskip
After placing the patterns, we had to orient them perpendicularly to the local gradient of the image. To do so we just store for each image the local Sobel gradient following the $x$ axis and the local Sobel gradient following the $y$ axis. Then we estimate the local angle of the gradient with the following formula:

\begin{equation*}
\theta(x,y) = \arctan\left(\frac{\Delta x}{\Delta y}\right)
\end{equation*}

\begin{figure}
\centering
\includegraphics[width = 0.38\textwidth]{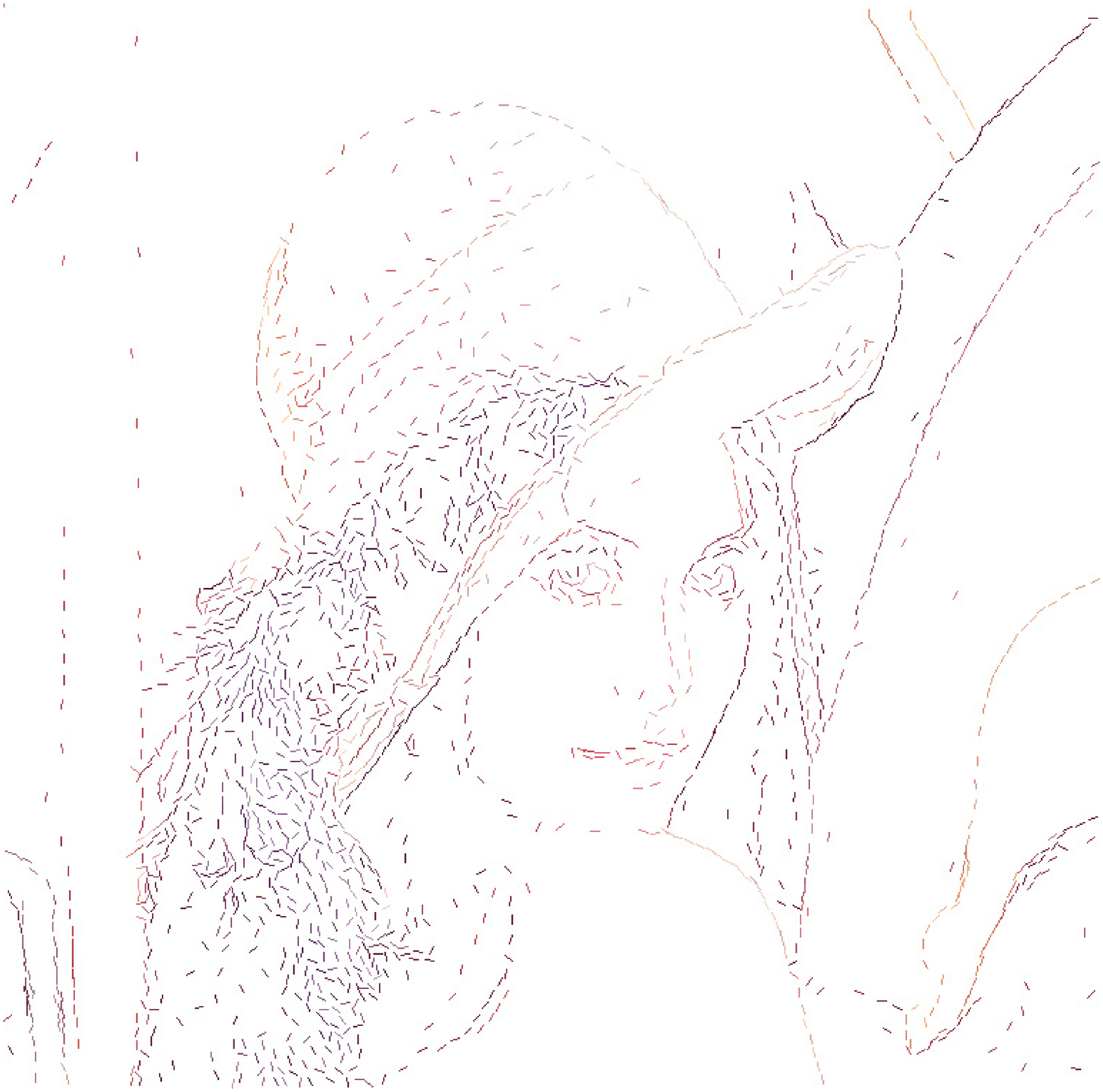}
\caption{Use of patterns to describe sharp edges in the image --- 2000 segments.}
\label{patternlena}
\end{figure}

Once this operation is done we can associate each frequency Site with its orientation. We obtain on the {\tt Lena} image the result shown on the  figure~\ref{patternlena}.

\medskip
Contrary to the previous frequency approach, we need less point to obtain a pleasant render. Thus we increase the rapidity of our algorithm and save some place on the hard drive. To get a good rendering, we need to take in consideration the scale of these shapes into account in order to avoid the collision between them. 
This is possible by calculating an individual scaling factor for each of them.

\section{Experiments and results}
\subsection{Blue noise characteristics}

\begin{figure}
\centering
\includegraphics[width=0.40\textwidth]{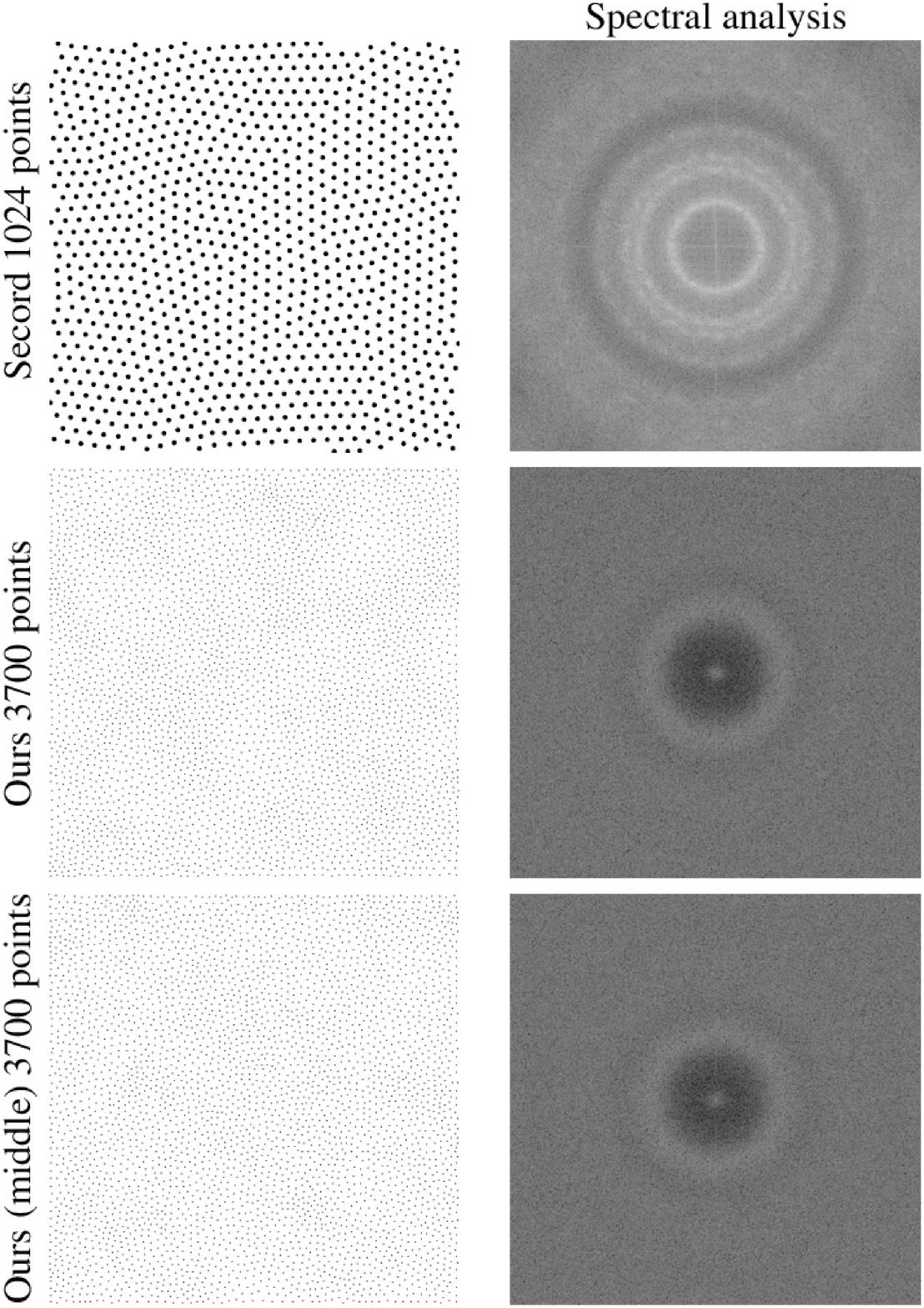}
\caption{Lloyd's method generates point distributions with regular structures if it is not stopped manually. The example set of 1024 points was computed with Lloyd's method to full convergence and contains regularities. The two other examples have been generated by our algorithm, the first directly and the second after the computation of a whole video as the last frame of this video. They present less regularities.}
\label{fft}
\end{figure}

Figure~\ref{fft} shows a representative distribution of 1024 random points that was generated with Lloyd's method and extracted form the paper of Balzer et al.~\cite{DBLP:journals/tog/BalzerSD09}. The distribution clearly exhibits regularities that are underlined by the FFT analysis on the right of the figure. Our method generates a random set of points (here 3700 points) with less regularities. The results is substantially better and is conserved during a whole video. Two sets where generated and shown on figure~\ref{fft}. The first has been made on a single image, with a direct point placement with our algorithm and present good characteristics. The second has been build inside a video sequence by successive point addition frame by frame until the final number of 3700 points. We notice that this distribution has the same characteristics as the previous one.

\medskip
These blue noise characteristics are interesting because generating a good random point set is a challenge. But we can't explain for now the reason of those results in our algorithm.

\subsection{Algorithm's complexity}

Our method and software allows one to convert a video into a stippling video, namely a time-varying point set with dynamic color and size attributes. 
The complexity of our method only depends on the number of support/site points that we used to stipple the video. 
This number of points depends in turn on the following user input entries: 
\begin{itemize}
\item The number of sites;
\item The factor for calculating the number of support points --- denoted by $\alpha$ previously.
\end{itemize}
The complexity of the algorithm is quadratic, $O(n^2)$ where $n$ is the number of sites. 
We have carried out several time measurements to confirm this overall complexity, and in order to identify some variations depending on other parameters. We have done the timing measures on a video containing 10 frames. 
The video represents a black disk whose size decreases progressively itself $10\%$ each frame.

\begin{figure}
\centering
\includegraphics[width=0.48\textwidth]{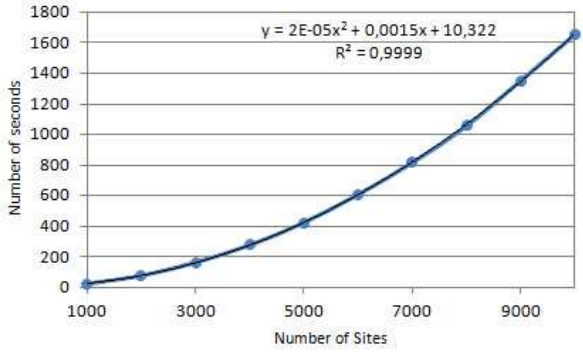}
\caption{Graph representing the time required to calculate a 10-frame video with varying number of sites.}
\label{graph1}
\end{figure}

\medskip
The timing statistics of that experiment are listed in Table~\ref{table:timing}. 
Each row stands for a frame (there are 10 frames), and each column denotes a number of point (from 1000 to 10000). 
All timings are expressed in seconds. 
The graph on the Figure~\ref{graph1} plot the amount of time consumed to stipple a video depending on the number required sites.

\medskip

\begin{table}
\centering
{\footnotesize
\begin{tabular}{| >{\columncolor{grisclair}} c|c|c|c|c|c|}
\hline \rowcolor{grisclair} & 1000 pts & 2000 pts & 3000 pts & 4000 pts & 5000 pts\\
\hline 0 & 9.5 & 35.0 & 76.5 & 134.7 & 208.6\\
\hline 1 & 4.8 & 14.1 & 32.3 & 49.6 & 75.5\\
\hline 2 & 3.1 & 8.9 & 18.1 & 30.7 & 47.5\\
\hline 3 & 2.4 & 6.1 & 12.3 & 21.4 & 32.2\\
\hline 4 & 1.7 & 4.4 & 8.5 & 14.2 & 21.3\\
\hline 5 & 1.3 & 3.2 & 6.0 & 9.8 & 14.6\\
\hline 6 & 1.1 & 2.4 & 4.3 & 6.8 & 9.9\\
\hline 7 & 0.9 & 1.8 & 3.1 & 4.8 & 7.0\\
\hline 8 & 0.8 & 1.4 & 2.3 & 3.5 & 5.0\\
\hline 9 & 0.7 & 1.4 & 1.8 & 2.6 & 3.5 \\

\hline \rowcolor{grisclair} & 6000 pts & 7000 pts & 8000 pts & 9000 pts & 10000 pts \\
\hline 0 & 299.6 & 409.6 & 529.4 & 670.0 & 832.4\\
\hline 1 & 112.9 & 144.6 & 187.6 & 259.1 & 302.2\\
\hline 2 & 5.8 & 89.4 & 115.6 & 145.7 & 180.6\\
\hline 3 & 45.8 & 59.5 & 76.7 & 97.0 & 118.5\\
\hline 4 & 30.1 & 40.7 & 52.1 & 65.2 & 80.6\\
\hline 5 & 20.2 & 27.1 & 37.6 & 43.5 & 52.8\\
\hline 6 & 13.7 & 18.5 & 23.7 & 29.3 & 35.9\\
\hline 7 & 9.5 & 12.7 & 16.1 & 19.9 & 23.9\\
\hline 8 & 6.7 & 8.7 & 11.1 & 13.6 & 16.3\\
\hline 9 & 4.7 & 6.1 & 7.7 & 9.3 & 11.1 \\
\hline
\end{tabular} 
\normalsize
}

\caption{Timing experiments for a toy sequence of $10$ frames --- in seconds.\label{table:timing}}
\end{table}

\medskip

During the first set of measures, we used a video with a lot of point suppression. 
We noticed that if we compute the same video backward (so the disk is getting bigger each frame) for the same number of site, the time needed was different!
In fact, we observed empirically that the addition of points requires more time than the operation of suppression.
 For $3000$ sites required on the frame with the biggest density, we needed $165$ seconds to stipple the video and $602$ seconds to compute the same video backward (the one with a lot of addition of points). 
Thus we emphasize on the fact that the stippling operation is not symmetrical although it visually leads to the same final result. 

\medskip
To estimate the time required to compute the video we need to take into account the number of addition of points during the shot. 
This operation increases the overall computation time.
Of course, one can accelerate our algorithm by porting it on graphics processor units (GPUs). 
The calculation of Voronoi diagrams is known to adapt well on GPU, and is far quicker and more efficient~\cite{DBLP:conf/isvc/VasconcelosSCG08} than CPU-based algorithms.

\subsection{Size of the output file}

\begin{figure}
\centering
\includegraphics[width=0.48\textwidth]{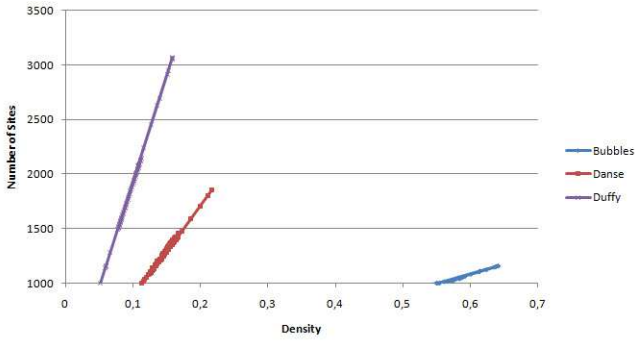}
\caption{Evolution of the number of Site by frame in function of the density.}
\label{pointstab}
\end{figure}

We measured for 3 different video shots of 91 frames the number of Sites depending on the frame density. The result is summed up on figure~\ref{pointstab}.

\medskip
We noticed that algorithm maintains a perfect correlation between those to parameters following the initial ratio asked by the user. This observation implies that we can estimate the average number of Point needed to compute each frame. However, in order to estimate the total number of Sites needed --- how many Sites are created during the computation --- we have first to know all the difference images and their density. Then we can estimate the total number of Sites with the formula:

\begin{equation*}
\frac{N_{\textrm{initial}}}{d_{\textrm{initial}}} + \sum_{i=0}^{\textrm{Number of frame}} \left[d_{\textrm{diff}}(i)*\frac{N_{\textrm{initial}}}{d_{\textrm{initial}}}\right]
\end{equation*}

Where $N$ represent the number of Sites and $d$ the density.

\medskip
With this method we can quickly estimate the final size of the output video file. After a ZIP compression we observed that we need 200 bites per frame per 1000 Sites. Thus for a stipple video with 25 frames per second, that lasts 1 hour, we encode a video in about 20 Mb.

\medskip
An possible way to improve the compression of the output file would be to store for each frame the characteristics of the points that are added and those of the suppressed point. For example for a video with 2000 Sites per frame and 3000 different Sites used for the whole video we would need to store the characteristics of 3000 Sites, whereas our current method needs $2000*N_{\textrm{Frame}}$ Sites. This improvement implies to consider that the Sites generated by our method are totally stable. In fact, they can move slightly during the shot to adapt with the variation of local intensity. But if we compute each with numerous iteration to reach a great stability, this movement is really small. So suppressing all the movement of point would result only on the complete removal of what we could see as the flickering effect. Indeed in our current video output the little jittering effect results of this movement and of a to big threshold taken for judging of the stability of the Site during the computation. This choice has been made to improve the rapidity of the computation of the video.

\subsection{Primary colours}

\begin{figure}
\centering
\includegraphics[width=0.38\textwidth]{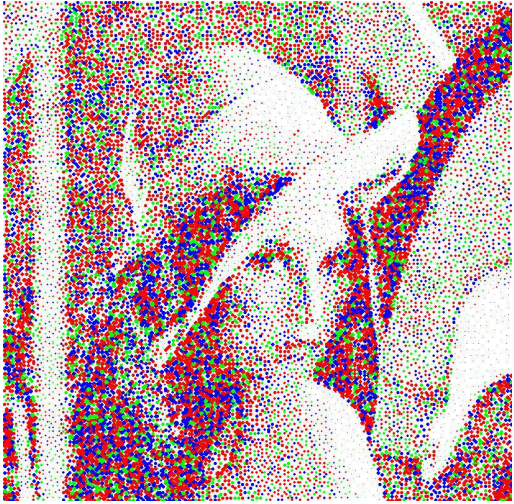}
\caption{Stippling on {\tt Lena} with primary colours. (15000 points)}
\label{lenaprimary}
\end{figure}

In order to be able to produce a render closer to some paintings we implements an algorithm that produces stippling with only primary coloured Sites --- red, green and blue Sites. This extension is quite simple for a single image: we just convert each colour in red green or blue with a probability equal to its proportion of red, green or blue. An example is shown on {\tt Lena} on figure~\ref{lenaprimary}. But it raises some problems for video:

\begin{itemize}
\item We have to maintain a high density of Sites. The eye makes a mean of the different colours an reconstruct the appropriate colour for each part of the image. That's why it is important to have a very high density of small points;
\item Having to maintain a high density of Sites increases the calculation time --- proportional to $n^2$ where $n$ is the number of Sites;
\item It increases the memory space needed to store the output video;
\item It increases the time needed to read each frame of the video and slower the frame rate in the viewer.
\end{itemize}

\medskip
Even if this approach generates drawings with only three colours, we looses some of the advantages of the Stippling process such as few colour transition, few storage space needed, few memory needed to read the video... To have a nice render we finally need to have a very high resolution of the stippling image and this put the stipple image with a number of Sites of the same order or about 10 times fewer as the number of pixel of the original. For example the image on the figure~\ref{lenaprimary} is drawn with 15000 Sites whereas the original image was composed of $256*256=65536$ pixels, wich is only 4 times higher.

\section{Discussion and concluding remarks}

Stippling is an engaging art relating to non-photo-realistic depiction of images and videos by using point sets rendered with various color and size attributes. 
Besides the pure artistic and aesthetic interests of producing such renderings, the stippling process finds also many other advantages in its own: 

\begin{itemize}
\item
The output video is fully vectorized, both on the spatial and temporal axes. 
Users can  interactively rescale video to fit the device screen resolution and upscale the frame rate as wished for fluid animations. 
Stippling could thus be useful for web designers that have storage capacity constraints and yet would like to provide video contents that yields the same appearance on various types of screen resolution devices (let it be PDA, laptop or TV). 
Furthermore, once the stippling process has been carried out, users can still personalized the rendering by tuning on-the-fly the size and other attributes of the points, without loosing much of the original semantic of the media.  

\item
Another characteristic of the stippling process is the production of a video that bear  only a few pixel color transitions compared to the original medium. 
Indeed, on a usual video, we have $\textrm{Width} \times \textrm{Height} \times N_{\textrm{Frame}}$ color transitions while 
in our output stippled video, we only have of number of transitions that is roughly $N_{\textrm{Point}}\times N_{\textrm{Frame}}$. 
This is advantageous in terms of energy savings for e-book readers for instance.
Stippling may potentially significantly increase battery life of such devices based on e-inks that consume energy only when flipping colors. 

\item
To improve the stippling process and correct the problem of  intensity loss of the stippled images (due to the averaging of the Human eyes), we can consider  the area of each Voronoi cell and use this measure to extract a multiplicative normalization factor for the site of the corresponding Voronoi cell. With this normalization factor, we are able to nicely correct the loss of intensity of the stippled images; 
The smaller a Voronoi cell, the bigger the normalization factor, and the darker the color setting.

\end{itemize}

\begin{figure*}
\centering
\includegraphics[width=\textwidth]{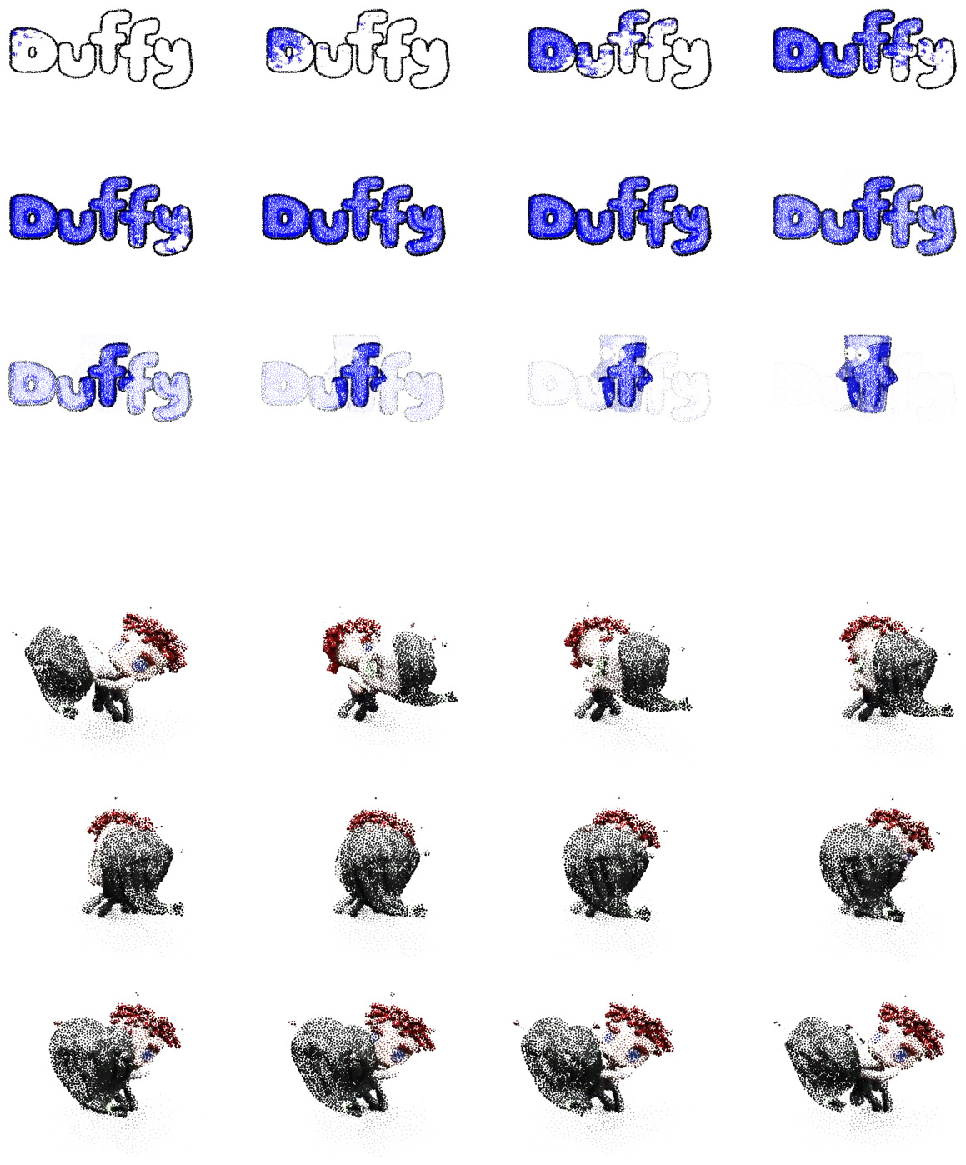}
\caption{These snapshots were extracted from two different output videos. See accompanying video. \footnotesize{This second particular sequence does not take into account the frequency information of images.}}
\label{screenshot}
\end{figure*}

Figure~\ref{screenshot} presents some extracted images from two stippled videos.
The accompanying video (in mp4 format, play with Quicktime please) illustrates various steps of our algorithm, and describe results on various video.
(An open source web applet allows one to play various stippled videos.)

\bibliographystyle{acmsiggraph}
\nocite{*}
\bibliography{biblio}

\end{document}